\documentclass[twocolumn,aps,superscriptaddress]{revtex4}
\usepackage{graphicx,amssymb,bm,natbib,amsfonts,amsmath,graphics,color}
\usepackage{bm}
\usepackage{hyperref}

\begin{document}

\title{Permanent tuning of quantum dot transitions to degenerate microcavity resonances}

\author{Jan Gudat}
\affiliation{Huygens Laboratory, Leiden University, P.O. Box 9504, 2300 RA Leiden, the Netherlands}

\author{Cristian Bonato}
\affiliation{Huygens Laboratory, Leiden University, P.O. Box 9504, 2300 RA Leiden, the Netherlands}

\author{Evert van Nieuwenburg}
\affiliation{Huygens Laboratory, Leiden University, P.O. Box 9504, 2300 RA Leiden, the Netherlands}

\author{Susanna Thon}
\affiliation{University of California Santa Barbara, Santa Barbara, California 93106, USA}

\author{Hyochul Kim}
\affiliation{University of California Santa Barbara, Santa Barbara, California 93106, USA}

\author{Pierre M. Petroff}
\affiliation{University of California Santa Barbara, Santa Barbara, California 93106, USA}

\author{Martin P. van Exter}
\affiliation{Huygens Laboratory, Leiden University, P.O. Box 9504, 2300 RA Leiden, the Netherlands}

\author{Dirk Bouwmeester}
\affiliation{Huygens Laboratory, Leiden University, P.O. Box 9504, 2300 RA Leiden, the Netherlands}
\affiliation{University of California Santa Barbara, Santa Barbara, California 93106, USA}

\begin{abstract}
We demonstrate a technique for achieving spectral resonance between a polarization-degenerate micropillar cavity mode and an embedded quantum dot transition. Our approach is based on a combination of isotropic and anisotropic tensile strain effected by laser-induced surface defects, thereby providing permanent tuning. Such a technique is a prerequisite for the implementation of scalable quantum information schemes based on solid-state cavity quantum electrodynamics. \end{abstract}

\maketitle
Single self-assembled quantum dots (QDs) embedded in microcavities are interesting systems for quantum information applications. Cavity-induced Purcell enhancement of the emitter spontaneous emission rate has been exploited to demonstrate efficient and reliable single photon sources \cite{lounisRPP05,  straufNP07,  forchelIOP10}. Moreover, quantum information schemes employing cavity quantum electrodynamics with quantum dots coupled to semiconductor microcavities have been proposed and implemented \cite{imamogluPRL99, reithmayerNature04, rakherPRL09, xuPRB09, bonatoPRL2010}. Such system would provide a scalable platform for hybrid quantum information protocols, in which photonic qubits are used for long-distance transmission and matter qubits for local storage and processing \cite{ciracPRL97, vanenkPRL97}.\\
Several quantum information applications require a polarization-degenerate cavity mode that is spectrally resonant with a specific quantum dot optical transition \cite{xuPRB09, bonatoPRL2010}. Polarization-degeneracy is needed in order to transfer an arbitrary polarization state of a photon to the spin of a single electron confined in the dot, or viceversa. In the case of micropillar cavities, due to residual strain in the structure or small shape asymmetries, the fundamental cavity mode often consists of two linearly-polarized submodes, energy split by an amount $\Delta E$. An important issue to note is the fact that the optical properties of a self-assembled quantum dot strongly depend on its specific size and local strain, neither of which is deterministically controllable in the growth process. Therefore post-fabrication tuning techniques are crucial to achieving exact spectral resonance.\\
The most flexible tuning technique is Stark-shifting: embedding the dots in a diode structure and applying a voltage leads to a shift of the optical transition frequency by the quantum confined Stark effect \cite{fryPRL2000}. Such shifts can be finely tuned to a limited range of a few hundred $\mu$eV, making the technique most effective in combination with some other coarse tuning procedures, such as temperature or strain. Temperature tuning, of either the whole sample \cite{kirazAPL01} or a local spot \cite{faraonAPL07} is an effective approach, with energy shifts on the order of $1-2$ meV reported in the literature. The temperature can, however, only be adjusted in the range of about $4-50$ K: at higher temperatures the dot luminescence quenches. Moreover, if one is interested in the spin of a single electron in the QD, it is crucial to keep the temperature below $30$ K, in order to avoid reducing the spin relaxation time \cite{paillardPRL01}.\\
Strain-tuning, via piezoelectric actuators or mechanical tips, has also been extensively investigated \cite{obermullerAPL99, zanderOE09, bryantPRL10}.  Recently, it was shown that strain control by means of laser-induced surface defects can be used to fine-tune the optical properties of semiconductor microcavities \cite{doornAPL96a, bonatoAPL2009}. By focusing a strong laser beam on a small spot, far away from the cavity center to preserve the optical quality of the device, the local birefringence can be modified. Here we show that, by a controlled combination of anisotropic and isotropic strain, one can simultaneously get a polarization-degenerate cavity and tune a dot optical transition into resonance with the cavity mode. Since the defects are permanent, no external tuning equipment is needed during an experiment, and this makes our technique ideal for scalability purposes.\\
\begin{figure}[ht]
\centering
\caption{\label{fig:micropillar} Sketch of the micropillar structure used in the experiments.}
\includegraphics[width=7 cm] {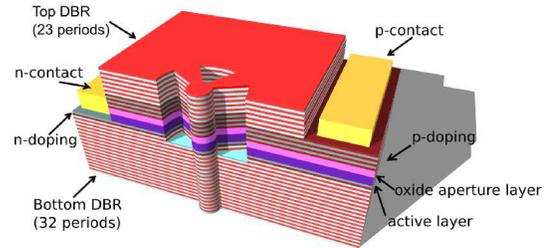}
\end{figure}
We investigated a sample with quantum dots embedded in micropillar cavities, grown by molecular-beam epitaxy on a GaAs [100] substrate. The microcavity consists of two distributed Bragg reflector (DBR) mirrors, made by alternating $\lambda/4$ layers of GaAs and Al$_{0.9}$Ga$_{0.1}$As. Between the mirrors, the active $\lambda$-GaAs layer contains embedded InGaAs/GaAs self-assembled quantum dots and sits underneath an AlAs oxidation layer. Trenches are etched down to the bottom DBR and the sample is placed in  asteam oven to define an AlO$_x$ oxidation front in the AlAs layer, providing transverse optical-mode confinement which results in high quality factors \cite{stoltzAPL05}. Using micropillars defined by trench shapes, intra-cavity electrical gating of multiple devices is possible by the fabrication of a PIN-diode structure (see Fig. 1 for a sample diagram).\\
Defects are created on the sample surface by a laser beam (about 100 mW/$\mu$m$^2$, $\lambda = 532$ nm) tightly focused on the structure for about 30 seconds by a high-NA aspheric lens $L_1$ (focal length $f_0 = 4.2$ mm, $NA = 0.6$). The material is locally melted and evaporated, leaving a hole which is approximately $ 2 \mu m$ wide and at least $2 \mu m$ deep. The whole process is performed in a helium-flow cryostat, at a temperature of $4K$.\\
\begin{figure}[ht]
\centering
\caption{\label{fig:polSplitting} Frequency splitting of the two orthogonally-polarized submodes of the fundamental cavity mode as a function of the burnt holes.}
\includegraphics[width=7 cm] {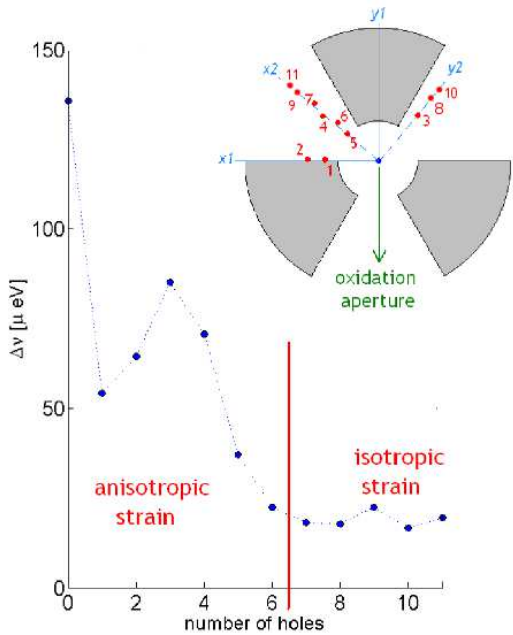}
\end{figure}
The first step consists of reducing the fundamental cavity mode to polarization-degeneracy, following the procedure described in \cite{bonatoAPL2009}. The built-in strain can be compensated by applying anisotropic strain, through holes burnt at proper positions. The direction of the original built-in strain is however unknown, so one must use a trial-and-error procedure, illustrated in Fig. \ref{fig:polSplitting}. We first start burning a hole at a random orientation, for example along the direction labeled in the figure as $x_1$. If the splitting gets larger,  we move to the orthogonal direction. If the splitting decreases, we keep burning holes until the splitting stops decreasing. In the example shown in Fig. \ref{fig:polSplitting}, the first hole reduces $\Delta E$ from $140 \pm 4$ $\mu$eV to $54 \pm 1$ $\mu$eV, but a second one slightly increases it. This is an indication that all the strain along that particular direction was compensated. We repeat the same procedure on a reference system rotated by $45$ degrees with respect to $[x_1, y_1]$. In the example, we start burning the third hole along $y_2$, which increases the splitting to $\Delta E = 82.6 \pm 0.4$ $\mu$eV. Therefore we switch to the orthogonal direction  $x_2$. Burning holes along this direction reduces $\Delta E$ to around $15$ $\mu$eV. The procedure can be further iterated along directions in between $x_1$ and $x_2$ and generally leads to splittings smaller than the mode linewidth (in our system about $50$ $\mu$eV), which is the requirement for quantum information experiments. No appreciable change in the cavity quality factor was observed.\\
\begin{figure}[ht]
\centering
\caption{\label{fig:vscan} Voltage-resolved photoluminescence plots for the holes described in Fig. 2. Originally, the cavity mode is non-degenerate (splitting around $140$ $\mu$eV) and  QD-3 is around $0.5$ meV detuned to the blue-side of the cavity mode. Burning $6$ holes reduces the splitting to about $15$ $\mu$eV and QD-3 is about $0.1$ meV detuned. Applying isotropic strain, by burning pair of holes along orthogonal direction, the dot can be brought into resonance with the cavity mode, without destroying the mode degeneracy (see plot for $11$ holes, bottom right). See Fig. 2 for the position of the holes.}
\includegraphics[width=7 cm] {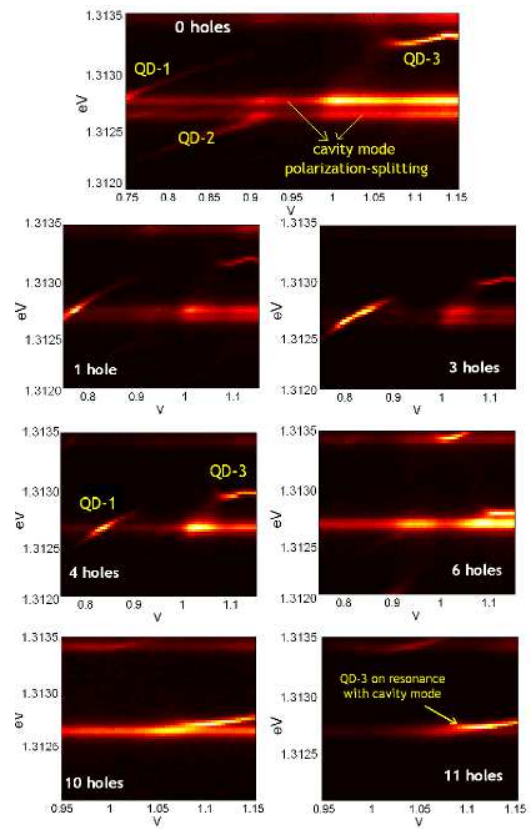}
\end{figure}
Strain affects the optical transitions of the quantum dots as well. In Fig. \ref{fig:vscan}, we show plots of voltage-resolved photoluminescence from the same microcavity analyzed in Fig. 2. We pump the sample non-resonantly with about 1 $\mu$W/$\mu$m$^2$ laser beam at $785$ nm, above the GaAs bandgap, and we spectrally resolve the photoluminescence with a spectrometer (resolution $25$ $\mu$eV/pixel). Scanning the voltage of the microstructure PIN-diode, different charged states of the dot can be selected \cite{warburtonNat00} and the frequency of the optical transitions can be tuned by the Stark effect \cite{fryPRL2000}. The flat lines in the plots correspond to the fixed frequency emission of the fundamental cavity mode, split into two orthogonally-polarized submodes. The Stark-shifting lines correspond to QD optical transitions.\\
The effect of laser-induced defects is always a red-shift of the optical transition, independent of the actual position of the hole. The shift of the dot transition is generally much larger than the corresponding shift of the cavity mode, and from a sample of more than one hundred holes burnt, the ratio of the shifts was found to be on average $5:1$. These findings can be explained with a simple model \cite{newPaper}. The fact that the optical transition always red-shifts suggests that by burning holes we effectively apply tensile strain to the structure. This could be explained by assuming that, by removing material, we release some compressive strain that pre-exists due to lattice-mismatch in the dot. Such tensile strain affects the band-structure both of the InAs dot material and of the bulk surrounding GaAs, reducing the InAs energy gap and the width of the confining potential well. The change in the band-structure profiles can be shown to be independent of the direction of the strain in the plane of the dot \cite{newPaper}.\\
The difference in the way the cavity mode and the dot transition are affected by hole-burning can be exploited to tune a quantum dot transition into resonance with a polarization-degenerate cavity. In Fig. \ref{fig:vscan} one can see that, while burning the first six holes, needed to reduce the splitting $\Delta E$, the optical transitions of the dots red-shift, so that the transitions labeled as QD-1 and QD-2, originally resonant with the non-degenerate fundamental cavity mode, tune out of resonance. After burning six holes we have a polarization-degenerate cavity mode, with a quantum dot transition (labeled QD-3) about $100$ $\mu$eV detuned on the blue-side. Now the challenge is to shift this transition into resonance, without perturbing the cavity mode degeneracy. This can be done by applying isotropic strain: we can burn sets of two holes at orthogonal directions, for example one along $x_2$ and the other along $y_2$, at the same distance from the center. This leaves the splitting $\Delta E$ unaltered while red-shifting the dot transition. The results are shown in the bottom two pictures in Fig. \ref{fig:vscan}, corresponding to the tenth and eleventh hole burnt. The dot is finally on resonance and the fundamental cavity mode splitting is $13 \pm 1$ $\mu$eV (right side of Fig. 2, for holes 7-11).\\
In conclusion, we demonstrated a tuning technique for micropillar cavities with embedded quantum dots, which allows us to obtain polarization-degenerate micropillars with a QD transition on resonance. Our technique is a crucial prerequisite for the implementation of scalable quantum information systems involving photon polarization and the spin of a single carrier trapped in the dot.\\

This work was supported by the NSF grant 0901886, the Marie-Curie award No. EXT-CT-2006-042580 and FOM$\backslash$NWO grant No. 09PR2721-2. We thank Andor for the CCD camera.

\end{document}